\documentclass[twocolumn,twoside,slac_two]{revtex4}
\usepackage{graphicx}
\usepackage{fancyhdr}
\begin{document}

\title{Radiative B Decays}

\author{D. Bard}
\affiliation{Imperial College, London, UK}

\begin{abstract}
I discuss recent results in radiative $B$ decays from the Belle and BaBar collaborations. 
I report new measurements of the decay rate and $CP$ asymmetries in $b \to s \gamma$ and $b \to d \gamma$ decays, and measurements of the photon spectrum in $b \to s \gamma$. 

\end{abstract}

\maketitle

\thispagestyle{fancy}

\section{Introduction}

\begin{figure*}[hbt]
\centering
\includegraphics[width=135mm]{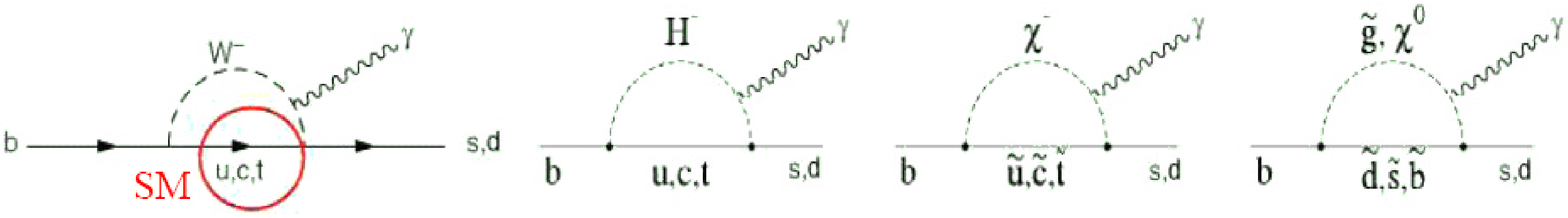}
\caption{Feynman diagrams for radiative penguin decays, showing the standard model loop and various new SUSY physics scenarios. } \label{loops}
\end{figure*}

Radiative penguin decays are flavour changing neutral currents which do not occur at tree level in the standard model (SM), but must proceed via one loop or higher order diagrams. 
These transitions are therefore suppressed in the SM, but offer access to poorly-known SM parameters and are also a sensitive probe of new physics. 
In the SM, the rate is dominated by the top quark contribution to the loop, but non-SM particles could also contribute with a size comparable to leading SM contributions (see Fig.~\ref{loops}). 
The new physics effects are potentially large which makes them theoretically very interesting, but due to  their small branching fractions they are typically experimentally challenging.

\section{\boldmath{$\lowercase{b} \to \lowercase{s} \gamma$} }
\begin{figure}[htb]
\centering
\includegraphics[width=80mm]{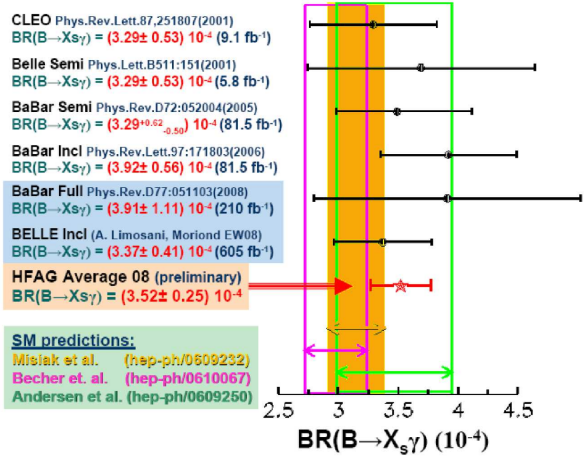}
\caption{Summary of theoretical predications and experimental results for the $b \to s \gamma$ branching fraction.  } \label{bsg-summ}
\end{figure}

Considerable work has gone into the theoretical prediction for the $b \to s \gamma$ branching fraction (BF), which has now been calculated at the next-to-next-to leading order as 
$(3.15 \pm 0.23) \times 10^{-4}$~\cite{misiak} and 
$(2.98 \pm 0.26) \times 10^{-4}$~\cite{becher}. 
A graphical comparison of theoretical predictions with experimental results is given in Fig.~\ref{bsg-summ}. 
There is currently good agreement between theory and experiment, and work continues to reduce the errors on measurements to even more stringently test the predictions.

As well as the branching fraction, quantities such as the photon energy spectrum and various asymmetries in $b \to s \gamma$ decays can be measured. 
The photon energy distribution depends on the mass ($m_b$) and Fermi motion ($\mu_b$) of the $b$ quark - the spectrum peaks at half $m_b$ as seen in Fig.~\ref{belle-bsg}. 
Measurements of the moments of the photon spectrum can be used to reduce the model-dependent error on the Cabibbo-Kobayashi-Maskawa (CKM) matrix elements $|V_{ub}|$ and $|V_{cb}|$. 

In the standard model the direct $CP$ asymmetry is less than 1\%, and new physics effects could enhance this to up to 15\%~\cite{hurth}. 
The $B^0B^+$ partial rate asymmetry, or isospin asymmetry, is predicted to be up to 10\%\ in the SM~\cite{kagan}.

\subsection{\boldmath{Recoil Method} }
A number of different experimental techniques exist to measure the $b \to s \gamma$ transition. 
They are optimised to reduce the significant continuum background ($e^+e^- \to q \overline{q}$ where $q = u, d, s, c$).

In the recoil method, one $B$ in the decay (the recoil or ``tag'' $B$) is fully reconstructed in a number of modes. 
The photon spectrum from other $B$ (the signal $B$) is measured. 
In a recent BaBar analysis~\cite{bsg-babar} using 210 $fb^{-1}$ of data over 1000 hadronic modes are used to reconstruct the tag $B$. 
The signal $B$ is constructed from one high-energy photon, plus all tracks and neutral particles not used in the reconstruction of the tag $B$. 
This technique allows the photon spectrum to be measured in the signal $B$ rest frame. 
The signal efficiency is low (around 0.3\%) but continuum background is almost eliminated. 
Using a photon energy cutoff of $E_{\gamma}>1.9$ GeV the $b \to s \gamma$ branching fraction was found to be: 
\begin{eqnarray}
BF(B \to X_s \gamma) = (3.65 \pm 0.85 \pm 0.60) \times 10^{-4} \nonumber \\
(E_{\gamma}>1.9\mathrm{\ GeV})\nonumber
\end{eqnarray}
where the first error is statistical and the second systematic. 
Extrapolating down to a photon energy limit of $E_{\gamma}>1.6$ GeV gives: 
\begin{eqnarray}
BF(B \to X_s \gamma) = (3.91 \pm 1.11) \times 10^{-4} \nonumber \\
(E_{\gamma}>1.6\mathrm{\ GeV}). \nonumber
\end{eqnarray}

\begin{figure}[htb]
\centering
\includegraphics[width=80mm]{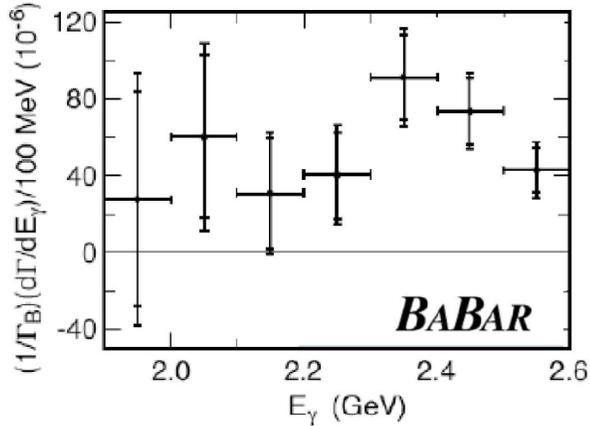}
\caption{Photon energy spectrum for $b \to s \gamma$ decays measured by the BaBar collaboration using a recoil method~\cite{bsg-babar}. } \label{babar-bsg}
\end{figure}

Using a more restricted photon spectrum of $E_{\gamma}>2.2$GeV the $CP$  and isospin asymmetries for $b \to (s,d) \gamma$ were found to be: 
\begin{eqnarray}
A_{CP} = 0.10 \pm 0.18 \pm 0.05\nonumber\\
\Delta_{0-} = -0.06 \pm 0.15 \pm 0.07\nonumber
\end{eqnarray}
respectively. 
From the photon spectrum, shown in Fig.~\ref{babar-bsg}, the $b$ quark mass and the Fermi energy can be calculated, giving: 
\begin{eqnarray}
m_b = 4.46 ^{+0.21}_{-0.23} \mathrm{\ GeV} \nonumber\\ \nonumber
\mu^2_{\pi} = 0.64 ^{+0.39}_{-0.38} \mathrm{\ GeV}^2. 
\end{eqnarray}

\subsection{Inclusive Method}
\begin{figure}[htb]
\centering
\includegraphics[width=70mm]{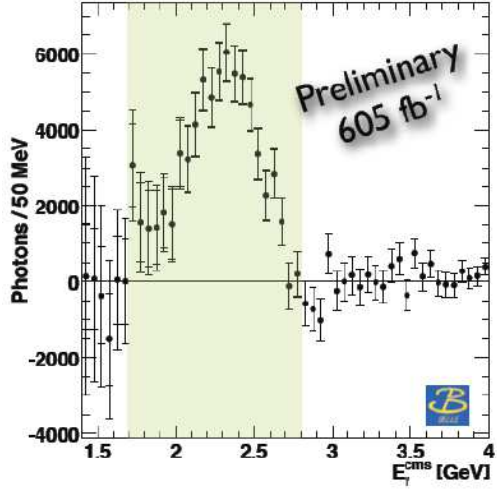}
\caption{The photon energy spectrum in $b \to s \gamma$ measured by the Belle collaboration using a fully inclusive method, after background subtraction. The shaded area is the region used for the measurements described in the text. The absence of events outside this region indicates that the backgrounds have been successfully subtracted~\cite{bsg-belle}. } \label{belle-bsg}
\end{figure}

The first measurement of $b \to s \gamma$ to extend down to a photon energy of $E_{\gamma}>1.7$GeV has been made by the Belle collaboration, using 605$fb^{-1}$ of data~\cite{bsg-belle}. 
The technique used is a fully inclusive one, where only the signal photon is reconstructed and the background from non-$B$ decays is reduced using lepton tags from the tag $B$. 
Vetoes are used to remove photons from $\pi^0$s and $\eta$s by rejecting high energy photons if, when paired with any other photon in the event, they possess an invariant mass near that of a $\pi^0$ or $\eta$.
Topological event information is used to suppress continuum backgrounds - $B\overline{B}$ decays tend to be spherical in shape in the centre-of-mass (CM) frame, whereas continuum events are more jet-like. 
After cuts have been made, there still remains some background events which are subtracted using off-resonance data for continuum background, and Monte Carlo simulated events for other $B$ backgrounds. 
Figure~\ref{belle-bsg} shows the photon energy spectrum after background subtraction. 

The branching fraction for $b \to s \gamma$ is found to be: 
\begin{eqnarray} \nonumber
BF(B \to X_s \gamma) = (3.31 \pm 0.19 \pm 0.37 \pm 0.01) \times 10^{-4}\\ \nonumber
 (E_{\gamma}>1.7\mathrm{\ GeV}) 
\end{eqnarray}
where the first error is statistical, the second systematic and the third due to uncertainty in the boost. 
Extrapolating to photon energies above $E_{\gamma}>1.6$ GeV gives: 
\begin{eqnarray} \nonumber
BF(B \to X_s \gamma) = (3.31 \pm 0.41) \times 10^{-4}\\
 (E_{\gamma}>1.6\mathrm{\ GeV}) \nonumber
\end{eqnarray}

From the photon energy spectrum, shown in Fig.~\ref{belle-bsg} the first and second moments are found to be
\begin{eqnarray}
<E_{\gamma}> = 2.281 \pm 0.032 \pm 0.053 \pm 0.002 \mathrm{\ GeV} \nonumber \\
<E_{\gamma}^2> - <E_{\gamma}>^2 = 0.0396 \pm 0.0214 \pm 0.0012 \mathrm{\ GeV}^2 \nonumber
\end{eqnarray}
respectively.

\subsection{\boldmath{Semi-inclusive Method} }
\begin{figure}[htb]
\centering
\includegraphics[width=60mm]{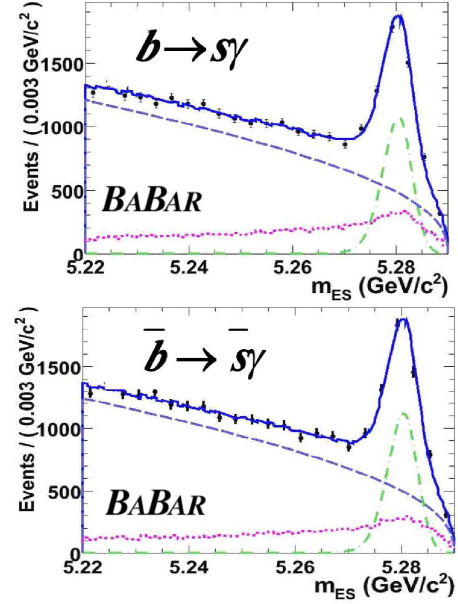}
\caption{$m_{ES}$ distribution for $b \to s \gamma$ decays (top) and $\overline{b} \to \overline{s} \gamma$ decays (bottom). 
Data is shown as points with error bars, the signal contribution in green dash-dotted line, $B\overline{B}$ in magenta dotted line, continuum background in blue dashed line ~\cite{bsg-babar-acp}. } \label{babar-acp}
\end{figure}

BaBar recently presented an updated measurement of the $CP$ asymmetry in $b \to s \gamma$ decays, made using a semi-inclusive method and 383$\times 10^6$ $B\overline{B}$ pairs~\cite{bsg-babar-acp}. 
In this type of analysis, the inclusive decay is approximated using a reconstruction of many exclusive final states. 
This analysis uses 16 exclusive $B \to X_s \gamma$ final states which cover approximately 50\%\ of the total width within the hadronic mass range of $0.6 < M(X_s)<2.8$GeV/c$^2$, which corresponds to a photon energy cutoff of $E_{\gamma}>1.9$ GeV. 
Continuum background is reduced by combining a number of event shape variables into a boosted decision tree, and fake high energy photons from $\pi^0$ or $\eta$ decays are removed using vetoes as described above. 
The $CP$ asymmetry is measured from a fit to the beam-constrained mass $m_{ES}$ in the $b \to s \gamma$ and $\overline{b} \to \overline{s} \gamma$ channels, as shown in Fig.~\ref{babar-acp}, where $m_{ES} = \sqrt{E^*_{beam} - p^*_B}$ (where $E^*_{beam}$ is the beam energy in the CM frame, and $P^*_B$ is the $B$ momentum in the CM frame). 
The result is the most accurate measurement to date of the direct $CP$ violation in this decay:
\begin{eqnarray} \nonumber
A_{CP}(b \to s \gamma) = -0.012 \pm 0.030 \pm 0.018,  \nonumber
\end{eqnarray}
where the first error is statistical and the second systematic. 
This is in good agreement with the standard model prediction of -1\%.

\section{\boldmath{$\lowercase{b} \to \lowercase{d} \gamma$} }
In the SM the rate for $b \to d \gamma$ is suppressed with respect to $b \to s \gamma$ by a factor of around 20, and is also particularly sensitive to new physics. 
Direct $CP$ asymmetries are expected to be $O$(10\%), but new physics effects could significantly enhance this. 
The ratio of CKM matrix elements $| V_{td}/V_{ts} |$ can be obtained from the ratio of the $b \to d \gamma$ and $b \to s \gamma$ BFs, for example in the ratio of the exclusive decays $B \to \rho \gamma$ and $B \to K^* \gamma$ using the formula~\cite{ali,ball}:
\begin{eqnarray}
\begin{small}
\frac{BF(B \to \rho \gamma)}{BF(B \to K^* \gamma)}
= \left|\frac{V_{td}}{V_{ts}}\right|^2 
\frac{(1-m_{\rho}^2 / m_B^2) ^3}{(1-m_{K^*}^2 / m_B^2) ^3}
\zeta^2[1+\Delta R]
\label{vtdvts}
\end{small}
\end{eqnarray}
where $\zeta$  is the ratio of form factors for $B \to \rho \gamma$ and $B \to K^* \gamma$ and $\Delta R$ is a factor to account for the differences in decay dynamics.

\subsection{\boldmath{$B \to (\rho,\omega) \gamma$ Branching Fractions and  $|V_{td}/V_{ts}|$} }
The exclusive $b \to d \gamma$ decays $B \to (\rho,\omega) \gamma$ have been extensively studied by both the Belle and BaBar collaborations. 
Background from continuum decays is more significant than for $b \to s \gamma$ analyses and contamination from mis-identified $b \to s \gamma$ decays is also a problem.

The BaBar analysis~\cite{babar-rho} uses a combination of lepton tags to suppress continuum background and constructs a neural net (NN) containing event shape variables which is used in the fit to a dataset of 347$\times 10^6 \ B\overline{B}$ pairs. 
Also included in the fit are $m_{ES}$, $\Delta E$ and the helicity angle, where $\Delta E$ is the energy difference between the beam and the reconstructed $B$ meson $\Delta E = E^*_{beam} - E^*_B$, where $E^*_B$ is the CM energy of the $B$. 
The Dalitz angle is also included in the fit to $B^0 \to \omega \gamma$. 
The measured branching fractions are: 
\begin{eqnarray}
BF(B \to \rho \gamma) = (1.36 \pm 0.28 \pm 0.10) \times 10^{-6} \nonumber \\ \nonumber
BF(B \to (\rho,\omega) \gamma) = (1.25 \pm 0.25 \pm 0.09) \times 10^{-6} 
\end{eqnarray}
where the first error is statistical and the second systematic. 

\begin{figure}[htb]
\centering
\includegraphics[width=70mm]{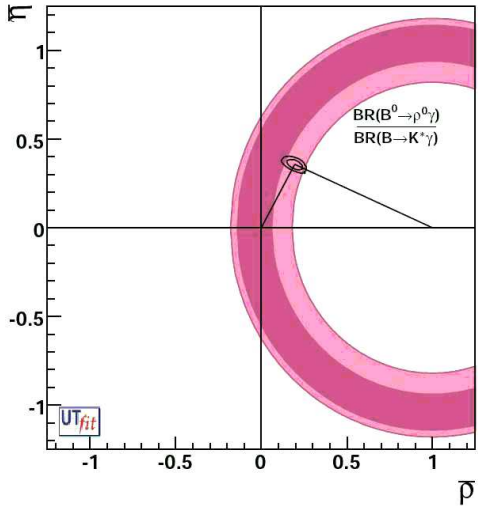}
\caption{The limits on $|V_{td} / V_{ts}|$ for the ratio of branching fractions of the neutral $\rho$ and K$^*$ decays, shown in the $( \overline{\rho}, \overline{\eta} )$ plane. } 
\label{vtdvts-rho0}
\end{figure}

\begin{figure}[htb]
\centering
\includegraphics[width=70mm]{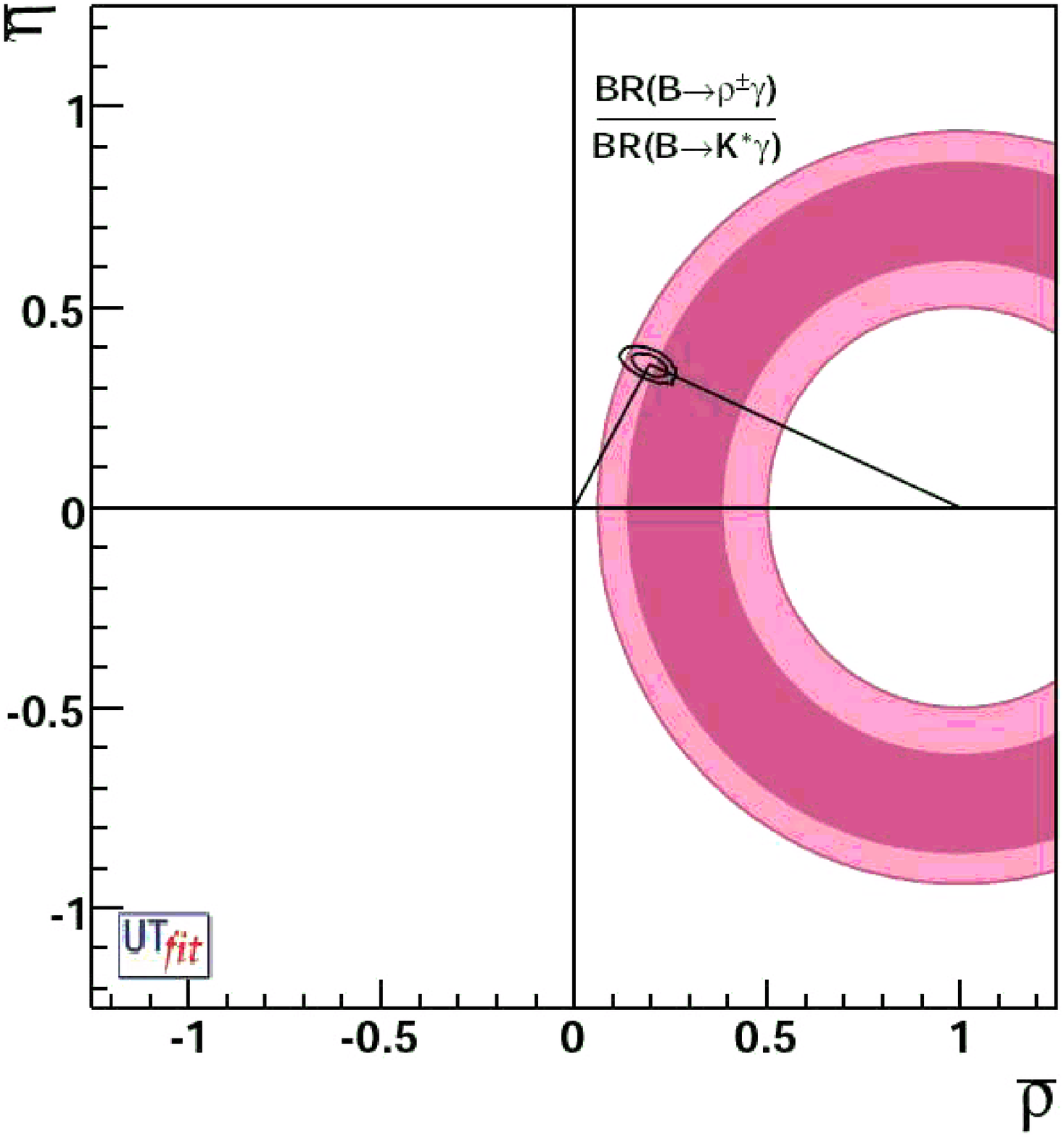}
\caption{The limits on $|V_{td} / V_{ts}|$ for the ratio of branching fractions of the charged $\rho$ and K$^*$ decays, shown in the $( \overline{\rho}, \overline{\eta} )$ plane. } 
\label{vtdvts-rhoCh}
\end{figure}

The equivalent Belle analysis~\cite{belle-rho} uses 657$\times 10^6 \ B\overline{B}$ pairs, and also deals specifically with the backgrounds from $B \to K^* \gamma$ decays. 
The fit uses $m_{ES}$, $\Delta E$ and, in the channel $B \to \rho^0 \gamma$, the invariant mass of the $\pi\pi$ pair with kaon mass assigned to one of the pions. 
The measured branching fractions are
\begin{eqnarray}
BF(B \to \rho \gamma) = (1.21 ^{+0.24}_{-0.22} \pm 0.12) \times 10^{-6} \nonumber \\
BF(B \to (\rho,\omega) \gamma) = (1.14 \pm 0.20 ^{+0.10}_{-0.12}) \times 10^{-6}. \nonumber
\end{eqnarray}
Both results from Belle and BaBar are in agreement with each other and with SM predictions. 
To extract $|V_{td} / V_{ts}|$ we use the formula given in~(\ref{vtdvts}) and  the theoretical quantities $\zeta = 0.85$ and $\Delta R = 0.1$. 
Using the world average branching fractions, we obtain:
\begin{eqnarray}
\left|\frac{V_{td}}{V_{ts}}\right| = 0.206 \pm 0.018,  \nonumber
\end{eqnarray}
in agreement with the limit obtained from $B_s / B_d$ oscillations~\cite{mixing}. 
This is represented graphically in the $( \overline{\rho}, \overline{\eta} )$ plane for $B^0 \to \rho^0 \gamma$ in Fig.~\ref{vtdvts-rho0} and for $B^+ \to \rho^+ \gamma$ in Fig.~\ref{vtdvts-rhoCh}.

\subsection{\boldmath{CP asymmetry in $B \to \rho \gamma$}}
Belle recently presented a measurement of the direct $CP$ asymmetry in $B^+ \to \rho^+ \gamma$ using $657\times 10^6\ B\overline{B}$ pairs~\cite{belle-rho-acp}. 
A simultaneous fit is performed to $m_{ES}$ and $\Delta E$ for $B^+ \to \rho^+ \gamma$ and $B^- \to \rho^- \gamma$ decays. 
Potential asymmetries from background sources are included in the systematic error, and a control sample of $B \to D \pi$ decays is used to understand bias in the detector. 
The direct $CP$ asymmetry is measured as: 
\begin{eqnarray}\nonumber
A_{CP}(B^+ \to \rho^+ \gamma) = 0.11 \pm 0.32 \pm 0.09, 
\end{eqnarray}
where the first error is statistical and the second systematic. 
The result is not statistically significant, but it agrees with standard model predictions of ~10\%.

\subsection{\boldmath{Inclusive $b \to d \gamma$} }
The first measurement of non-resonant $b \to d \gamma$ has been made by BaBar using 343$\times 10^6\ B\overline{B}$ pairs, performing a semi-inclusive analysis to approximate the inclusive decay~\cite{babar-bdg}. 
Seven exclusive final states were used, with up to four pions and up to one $\pi^0$ or $\eta$. 
The measurement was limited to the mass range $1.0 < M(X_d) < 1.8$ GeV/c$^2$ to exclude the $\rho$ and $\omega$ resonances and found 
\begin{eqnarray}
BF(B \to X_d \gamma) = (3.1 \pm 0.9 \pm 0.7) \times 10^{-6} \nonumber
\end{eqnarray}
in this mass range. 
Work is on-going to convert this to a fully inclusive measurement and a determination of $|V_{td} / V_{ts}|$. 

\section{Conclusion}
Measurements of the $b \to s \gamma$ decay are becoming ever more precise, in theory and experiment. 
Measurements have been made of the branching fractions with photon energy cutoffs at $E_{\gamma} < 1.7$ GeV and $E_{\gamma} < 1.9$ GeV, and $CP$ asymmetry in the mass range $0.6 < M(X_s)< 2.8$ GeV/c$^2$, showing no deviation from the current SM predictions. 

New measurements of $b \to d \gamma$ have also been made, with the experimental error on the branching fractions continuously decreasing and the first measurement of the $CP$ asymmetry in $B^+ \to \rho^+ \gamma$ being made. 
The first evidence for non-resonant $b \to d \gamma$ has been presented, and more interesting results are promised in the near future. 

Radiative penguin decays continue to be a rich source of information on little-known SM parameters and are unique probe into physics beyond the standard model.

\end{document}